\documentclass[12pt]{article}

\usepackage{tikz}

\usepackage{lineno}

\usepackage{amsmath}
\usepackage{graphicx}
\usepackage{enumerate}
\usepackage{natbib}
\usepackage{url} 

\newcommand{\blind}{0}

\newcommand{\fct}{\texttt}

\begin{document}

\def\spacingset#1{\renewcommand{\baselinestretch}%
{#1}\small\normalsize} \spacingset{1}


\if0\blind
{
  \title{\bf Teaching Design of Experiments using Hasse diagrams}
  \author{Hans-Michael Kaltenbach\hspace{.2cm}\\
    Department of Biosystems Science and Engineering, ETH Zurich\\
    \url{michael.kaltenbach@bsse.ethz.ch}\\
    {\small\url{https://orcid.org/0000-0003-2500-1221}}}
  \maketitle
} \fi

\if1\blind
{
  \bigskip
  \bigskip
  \bigskip
  \begin{center}
    {\LARGE\bf Teaching Design of Experiments using Hasse Diagrams}
\end{center}
  \medskip
} \fi

\bigskip

\begin{abstract} 
Hasse diagrams provide a principled means for visualizing the structure of statistical designs constructed by crossing and nesting of experimental factors. They have long been applied for automated construction of linear models and their associated linear subspaces for complex designs. Here, we argue that they could also provide a central component for planning and teaching introductory or service courses in experimental design.

Specifically, we show how Hasse diagrams allow constructing most elementary designs and finding many of their properties, such as degrees of freedom, error strata, experimental units and denominators for $F$-tests. Linear (mixed) models for analysis directly correspond to the diagrams, which facilitates both defining a model and specifying it in statistical software. We demonstrate how instructors can seamlessly use Hasse diagrams to construct designs by combining simple unit- and treatment structures, identify pseudo-replication, and discuss a design's randomization, unit-treatment versus treatment-treatment interactions, or complete confounding. These features commend Hasse diagrams as a powerful tool for unifying ideas and concepts.
\end{abstract}

\noindent%
\vfill

\newpage

\section{INTRODUCTION}
Most statistical designs encountered in introductory courses---such as  (fractional) factorial, randomized complete block, Latin square, and split-unit designs---can be created by crossing and nesting of factors. Hasse diagrams provide a visual representation of such designs~\citep{lohr2006} and have long been used in the statistical literature, in particular for discussing the analysis of orthogonal designs~\citep{tjur1984, Speed1987a, Andersson1990}, for defining algorithms to automate the analysis of a given design~\citep{Grossmann2014, Bate2016, Bate2016a, goos2012}, and to describe the linear subspaces of a design~\citep{Bailey2008}. However, they are used in only two textbooks~\citep{Oehlert2000,Bailey2008} as a supplementary tool for teaching experimental design, and have found no application as a central component to unify ideas and concepts for teaching.

Here, our goal is to reinforce the case made almost 25 years ago by \citet{lohr1995} that Hasse diagrams are ``useful in teaching others about design of experiments: because they are visual rather than analytic, they give students a supplementary perspective on different designs.'' In particular, we show how they provide instructors with a unifying tool to develop many of the classical designs by combining treatment and unit factors in different ways via crossing and nesting, and invite discussion of successful ``design patterns'' that can be combined to create more complex designs. This approach emphasizes finding a suitable design by looking at the relation between factors rather than relying on a table of known designs and picking the most suitable one. It is thus similar in spirit to the "no-name approach" by \citet{Lorenzen1993} and the focus on the data layout by \citet{Cobb2008}. In addition, the visual character of the diagrams encourages exploring different design alternatives. We use Hasse diagrams as a main teaching tool in a service course on introductory experimental design to biology / biotechnology MSc students in our department.

With \citet{Bailey2008}, we explicitly distinguish an experiment's \emph{unit structure} from its \emph{treatment structure}, use separate Hasse diagrams for the two structures, and create the diagram of an experimental design by connecting each treatment factor to the unit factor on which it is randomized. This idea has attracted surprisingly little attention in expositions of experimental design geared towards non-statisticians, an early exception being \citet{Cox1958}. It directly clarifies the concept of an experimental unit~\citep{hurlbert2013}, explicitly distinguishes between a unit-by-treatment from a treatment-by-treatment interaction~\citep{cox1984,berringtondegonzalez2007}, and shows that the same unit and treatment structures can be combined in different ways, leading to very different experimental designs. It also allows easier recognition of pseudo-replication~\citep{Hurlbert1984} and (inadvertent) split-unit designs~\citep{goos2012}. 

The emphasis on the factor structure of a design is complemented by the fact that a corresponding linear (mixed) model and its specification for a statistical software like \texttt{R}~\citep{RCoreTeam2019} can be derived directly from a design's diagram. This enables non-specialists to analyze data from a complex designed experiment once the Hasse diagrams have been constructed. The diagrams also allow extracting the ANOVA skeleton table, degrees of freedom for each experimental factor, and correct denominator factor for $F$-tests, which can all be used as sanity checks of the software output.

In the following, we always assume a balanced design with equal number of replicates. While Hasse diagrams can also describe unbalanced designs, most of the simple algorithms for calculating degrees of freedom or for deriving a linear model do not work in this case. We do not claim originality for any specific part of the following exposition; rather, we tried to collect and organize useful ideas for applying Hasse diagrams as a unifying component for teaching classical experimental designs, their properties, and the corresponding linear (mixed) models, and hope to encourage instructors to integrate them into their own courses. We provide generic diagrams for most designs that can easily be adapted to an instructor's preferred specific examples; to also provide a concrete example, we use Yates' oat varieties data-set for illustrating a blocked split-unit design~\citep{yates1935}.

\section{A FIRST EXAMPLE}
\label{sec:treatmentstructure}

\subsection{Constructing Hasse Diagrams}
To define an experiment design, we construct three Hasse diagrams for (i) the treatment structure describing the treatment factors and their relation; (ii) the unit structure describing the organization of the experimental material by unit factors and their relation; and (iii) the experiment structure resulting from randomizing each treatment factor on a unit factor. Each Hasse diagram is a visual representation of the partial order induced by nesting of factors and we represent each experimental factor by a vertex and use edges for representing the nesting relation.

A Hasse diagram always contains a top node \fct{M}, representing a one-level factor for a general mean, and every other factor is nested in \fct{M}. If  a factor \fct{Y} is nested in a factor \fct{X}, then each level of \fct{Y} occurs together with only one level of \fct{X}. We then draw \fct{Y} below \fct{X} and connect the two vertexes with an edge. The diagrams are thus ``read'' from top to bottom and a factor reached by following an edge corresponds to a more fine-grained partition of the experimental material into groups receiving the same treatment or belonging to the same block, for example. If two factors \fct{X} and \fct{Y} are crossed, then each level of \fct{X} occurs with each factor of \fct{Y} and we write \fct{X} and \fct{Y}  next to each other without a connecting edge. They provide independent partitions of the experimental material and give rise to an interaction factor \fct{X:Y} nested in both \fct{X} and \fct{Y}. Notationally, we denote treatment factors in bold and unit factors in italics, write random factors in parentheses, and assume a single response factor, which we mark by underlining. We often augment a diagram by denoting the number of factor levels by superscripts. 

\subsection{A Simple Factorial Design}
To illustrate the constructions, we consider a factorial design with two treatment factors \fct{A} and \fct{B} with $a$ and $b$ levels, respectively, in a balanced completely randomized design with $n$ experimental units per treatment combination and an experiment size of $N=n\cdot a\cdot b$. Extensions of this example to factorials with more than two treatment factors and designs with nested treatment factors are straightforward and we omit the details here.

\begin{figure}
\centering
\begin{tikzpicture}[
  var/.style={rectangle, draw=none, minimum size=5mm},
  node distance=1.5cm and 2cm,
  font=\small
  ]
    \node (ut) {$\mathbf{M}^1$};
    \node (at) [below left of=ut] {$\mathbf{A}^a$};
    \node (bt) [below right of=ut] {$\mathbf{B}^b$};
    \node (abt) [below right of=at] {$\mathbf{A:B}^{ab}$};
    \draw (ut.south) -- (at.north);
    \draw (ut.south) -- (bt.north);
    \draw (at.south) -- (abt.north);
    \draw (bt.south) -- (abt.north);
    \node (labelt) [above of=ut, node distance=1cm] {(a)};
  
    \node (uu) [right of=ut, node distance = 4cm] {$\mathit{M}^1$};
    \node (eu) [below of=uu] {$(\underline{\mathit{E}})^N$};
    \draw (uu.south) -- (eu.north);
    \node (labelu) [above of=uu, node distance=1cm] {(b)};
  
    \node (u) [right of=uu, node distance = 4cm] {$\mathbf{M}^1_1$};
    \node (a) [below left of=u] {$\mathbf{A}^a_{a-1}$};
    \node (b) [below right of=u] {$\mathbf{B}^b_{b-1}$};
    \node (ab) [below right of=a] {$\mathbf{A:B}^{ab}_{(a-1)(b-1)}$};
    \node (e) [below of=ab] {$(\underline{\mathit{E}})^{N}_{N-ab}$};
    \draw (u.south) -- (a.north);
    \draw (u.south) -- (b.north);
    \draw (a.south) -- (ab.north);
    \draw (b.south) -- (ab.north);
    \draw (ab.south) -- (e.north);
    \node (label) [above of=u, node distance=1cm] {(c)};

\end{tikzpicture}
\caption{(a): Treatment structure diagram. Hasse diagram of generic treatment structure with two crossed factors and their interaction. (b): Unit structure diagram. Simple unit structure with single unit factor. (c): Experiment structure diagram. Randomizing each treatment factor on the same unit factor yields a two-factor factorial in a completely randomized design.}
\label{fig:generic}
\end{figure}
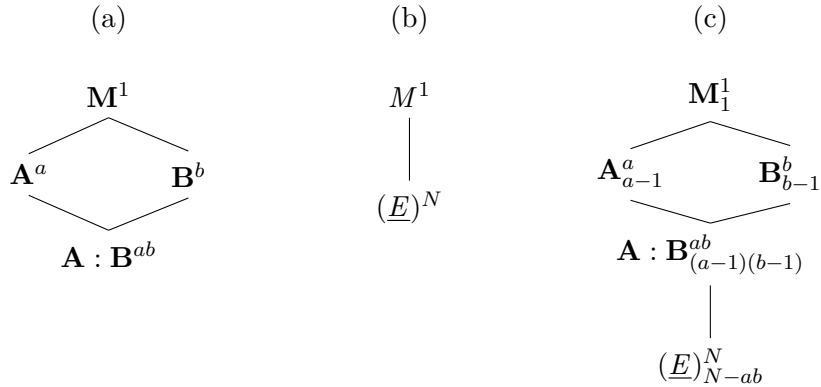

The treatment structure diagram for our example is given in Figure~\ref{fig:generic}(a). Our two treatment factors \fct{A} and \fct{B} are crossed such that each level of \fct{A} occurs with each level of \fct{B}, and both are nested in the general mean \fct{M}. The resulting interaction factor \fct{A:B} is nested in both \fct{A} and \fct{B}. 

The unit structure is shown in Figure~\ref{fig:generic}(b) and consists of a general mean and a single unit factor \fct{E} with $N$ levels. This factor is therefore the only possible experimental unit. It is good practice to use a descriptive name for this factor in a concrete example, e.g. \fct{Mouse}, \fct{Plot}, or \fct{Sample}. 

To construct the experiment design in Figure~\ref{fig:generic}(c), we first identify the two general means from the treatment and unit structures. We then draw an edge from each treatment factor to the unit factor on which it is randomized. This determines the experimental unit for each treatment factor, and nests it within its treatment factor(s). For our completely randomized design, we randomize both treatments \fct{A} and \fct{B} on the only available unit factor \fct{E}, which also makes it the experimental unit factor for \fct{A:B}. 

The diagrams give the transitive closure of the partial order on factors given by nesting. Consequently, we omit the direct edges from \fct{A} to \fct{E} and from \fct{B} to \fct{E}, since they are implied by nesting \fct{E} in \fct{A:B}, which in turn is nested in both \fct{A} and \fct{B}. We also remove the edge from \fct{M} to \fct{E} for the same reason. We might emphasize the balance in the design by writing $n\cdot a\cdot b$ instead of $N$ in the diagram, indicating that each treatment combination is replicated $n$ times.

\subsection{Degrees of Freedom}
\label{sec:df}
For any orthogonal design, we can calculate the degrees of freedom of a factor from the experiment diagram by subtracting from its number of levels the degrees of freedom of any factor in which it is nested~\citep{tjur1984,lohr2006}. Since the general mean \fct{M} has one level and is not nested in another factor, it has one degree of freedom. The interaction \fct{A:B} has $ab$ levels, from which we subtract one degree of freedom of \fct{M}, and $a-1$ respectively $b-1$ degrees of freedom of \fct{A} and \fct{B}. This yields the familiar $(a-1)(b-1)$ degrees of freedom. For the unit factor \fct{E}, we find $(n-1)ab$ degrees of freedom for a balanced design with $n$-fold replication and $N=nab$.

We may find that a factor \fct{Y} nested in \fct{X} has zero degrees of freedom. It is then completely confounded with \fct{X} and parameters and effects cannot be estimated individually. In our example, we might use a single experimental unit for each factor level combination (i.e., set $n=1$). The residual factor \fct{E} then has zero degrees of freedom and is completely confounded with the interaction \fct{A:B}. In other words, no replication exists for the treatment combinations and we can then only compute a pooled estimate of the combined variation of the interaction and the residual effects. This corresponds to merging \fct{A:B} and \fct{E} into a single factor in the diagram.

Another example is a one-way ANOVA design as shown in Figure~\ref{fig:marginality}(d) with only a single treatment factor level. The treatment is then confounded with the general mean, and we essentially have a simple sampling design where all units are assigned the same treatment and their responses are recorded.

\subsection{ANOVA Table, Error Terms, and $F$-Tests}
Using the Hasse diagram for the experiment design, we readily find the skeleton ANOVA table for an analysis. It contains one line for each factor in the diagram, where the general mean is excluded as usual and the lowest unit factor provides the residuals. For each line, the degrees of freedom are determined using the algorithm in Sect.~\ref{sec:df}, and the sum of squares and expected mean squares can be determined using similar algorithms, see for example \citet{lohr2006}. Each random unit factor gives rise to an error stratum in the table, and treatment factors randomized on this unit factor are tested within that stratum. 

The diagram facilitates identification of the error mean squares for the denominator of an $F$-test: it is provided by the closest random factor below the factor tested that does not involve a new fixed factor. If this factor is not unique, then no exact $F$-test exists. In our example, \fct{E} is the random factor nearest to each treatment factor and the $F$-test for testing the \fct{A} main effect, for example, is based on the ratio $\text{MS}(A)\,/\,\text{MS}(E)$ with $a-1$ numerator and $(n-1)ab$ denominator degrees of freedom.

\subsection{Linear Model}
Each factor provides a set of parameters for the linear model, with one parameter per factor level. For our example, the general mean is associated with a single parameter $\mu$, the treatment factors \fct{A}, \fct{B}, and \fct{A:B} have $a$ parameters $\alpha_i$, $b$ parameters $\beta_j$, and $ab$ parameters $(\alpha\beta)_{ij}$, respectively. The (random) residual unit factor \fct{E} has $nab$ levels corresponding to i.i.d.\ random variables $e_{ijk}$ with $k=1\dots n$; the corresponding population parameter is the variance component $\sigma^2=\text{Var}(e_{ijk})$. The full linear model for the example is thus
\begin{equation}
\label{eq:model}
  y_{ijk} \;=\; \mu+\alpha_i+\beta_j+(\alpha\beta)_{ij}+e_{ijk}\;,
\end{equation}
where $i=1\dots a$, $j=1\dots b$, and $k=1\dots n$. 

This model is over-parametrized, and each factor's subscript denotes how many of its parameters can be independently estimated. For example, \fct{A} is nested in \fct{M} and its parameters $\alpha_i$ are thus deviations from the general mean $\mu$ with expected response $\mu+\alpha_i$ for level $i$ of \fct{A}. If $\mu$ is given, we can only estimate $a-1$ out of the $a$ parameters of \fct{A}. This problem is resolved by imposing one additional relation either between the parameters or between their estimators, and \citet{nelder1994} provides a pointed discussion of the difference. With $ab$ fixed effect parameters, the variance $\sigma^2$ is estimated based on $(n-1)ab$ degrees of freedom. For $n=1$, merging \fct{A:B} with \fct{E} yields the additive model $y_{ij}=\mu+\alpha_i+\beta_j+e^*_{ij}$ where $e^*_{ij}=(\alpha\beta)_{ij}+e_{ij}$ with assumed $(\alpha\beta)_{ij}= 0$ and $k\equiv 1$ omitted.

\subsection{Symbolic Model Specification}
We can specify a linear model symbolically using the formulae proposed by \citet{wilkinson1973} and extended by \citet{tjur1984}. This model description uses the general operator \texttt{+} to add factors and \texttt{:} to denote interaction.  Common patterns are abbreviated, such as \texttt{A*B=A+B+A:B} for two fully crossed factors, or \texttt{A/B=A+A:B} for \texttt{B} is nested in \texttt{A}. The error strata are defined by the random unit factors, and the corresponding specification is written by \citet{tjur1984} using brackets: \texttt{A+[Blk]} describes a single treatment factor \fct{A} crossed with a random blocking factor \fct{Blk}. 

Formulating the model is then often a matter of writing out the diagram. For the example model~(\ref{eq:model}), the symbolic description is \fct{1+A+B+A:B} (abbreviated \fct{A*B}). It is equivalent to \fct{A/(A:B)+B/(A:B)} and to \fct{(A+B)/(A:B)}, which also describe the diagram. The additive model without interaction is \fct{A+B}.

We use the almost identical model specification for \texttt{R} for our examples. The general mean is then denoted \texttt{1}, but often omitted and then added implicitly. Also omitted is the lowest unit factor, corresponding to the lowest set of residuals. Error strata given by \texttt{[$\cdots$]} are implemented by the \texttt{Error($\cdots$)} part of a model specification.

\subsection{Marginality Principle}
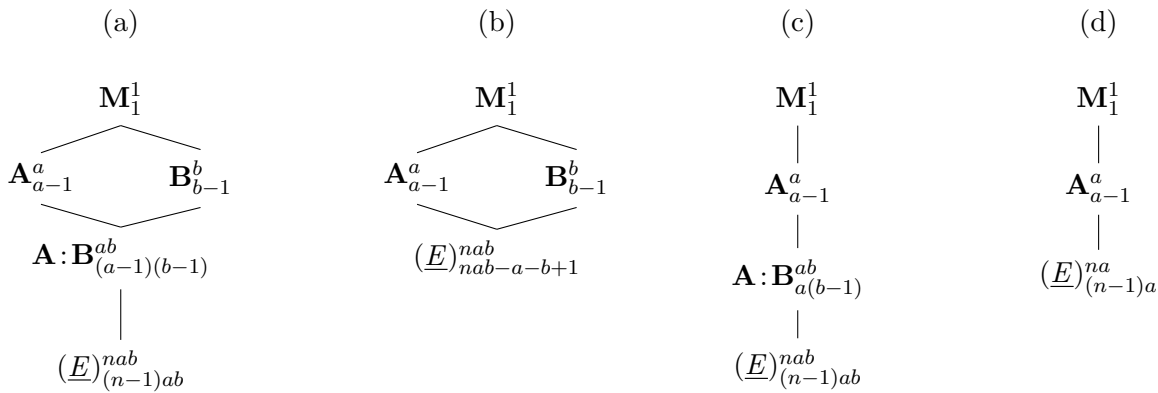
\begin{figure}
\centering
\begin{tikzpicture}[
  var/.style={rectangle, draw=none, minimum size=5mm},
  node distance=1.5cm and 2cm,
  font=\small
  ]
    \node (u) {$\mathbf{M}^1_1$};
    \node (a) [below left of=u] {$\mathbf{A}^a_{a-1}$};
    \node (b) [below right of=u] {$\mathbf{B}^b_{b-1}$};
    \node (ab) [below right of=a] {$\mathbf{A\!:\!B}^{ab}_{(a-1)(b-1)}$};
    \node (e) [below of=ab] {$(\underline{\mathit{E}})^{nab}_{(n-1)ab}$};
    \draw (u.south) -- (a.north);
    \draw (u.south) -- (b.north);
    \draw (a.south) -- (ab.north);
    \draw (b.south) -- (ab.north);
    \draw (ab.south) -- (e.north);
    \node (label) [above of=u, node distance=1cm] {(a)};
  
    \node (u1) [right of=u, node distance = 5cm] {$\mathbf{M}^1_1$};
    \node (a1) [below left of=u1] {$\mathbf{A}^a_{a-1}$};
    \node (b1) [below right of=u1] {$\mathbf{B}^b_{b-1}$};
    \node (e1) [below right of=a1] {$(\underline{\mathit{E}})^{nab}_{nab-a-b+1}$};
    \draw (u1.south) -- (a1.north);
    \draw (u1.south) -- (b1.north);
    \draw (a1.south) -- (e1.north);
    \draw (b1.south) -- (e1.north);
    \node (label1) [above of=u1, node distance=1cm] {(b)};
    
    \node (u2) [right of=u1, node distance = 4cm] {$\mathbf{M}^1_1$};
    \node (a2) [below of=u2, node distance=1.2cm] {$\mathbf{A}^a_{a-1}$};
    \node (ab2) [below of=a2, node distance=1.2cm] {$\mathbf{A\!:\!B}^{ab}_{a(b-1)}$};
    \node (e2) [below of=ab2, node distance=1.2cm] {$(\underline{\mathit{E}})^{nab}_{(n-1)ab}$};
    \draw (u2.south) -- (a2.north);
    \draw (a2.south) -- (ab2.north);
    \draw (ab2.south) -- (e2.north);
    \node (label2) [above of=u2, node distance=1cm] {(c)};
        
    \node (u3) [right of=u2, node distance = 4cm] {$\mathbf{M}^1_1$};
    \node (a3) [below of=u3, node distance=1.2cm] {$\mathbf{A}^a_{a-1}$};
    \node (e3) [below of=a3, node distance=1.2cm] {$(\underline{\mathit{E}})^{na}_{(n-1)a}$};
    \draw (u3.south) -- (a3.north);
    \draw (a3.south) -- (e3.north);
    \node (label3) [above of=u3, node distance=1cm] {(d)};
    
\end{tikzpicture}
\caption{Marginality principle. (a): the full model consists of two crossed factors and their interaction. (b): removing the interaction is compatible with the marginality principle. (c): removing the \fct{B} main effect and keeping the interaction violates the principle and leads to a design with \fct{B} nested in \fct{A}. (d): removing both the \fct{B} main effect and the interaction restores marginality.}
\label{fig:marginality}
\end{figure}

With only rare exceptions, the removal of a factor from a model should adhere to the \emph{marginality principle} and is admissible only when all higher-order interactions containing this factor are also removed~\citep{nelder1994}. In our example of a two-factor treatment design (Fig.~\ref{fig:marginality}(a)), removal of the interaction term \fct{A:B} observes the marginality principle (Fig.~\ref{fig:marginality}(b)). 

In contrast, removal of the main treatment factor \fct{B} while retaining the interaction \fct{A:B} violates this principle. The factors \fct{A} and \fct{B} are no longer crossed as intended by the treatment structure, but \fct{B} now appears nested in \fct{A} in the diagram (Fig.~\ref{fig:marginality}(c)).

Removal of the interaction factor together with one main effect factor (e.g., \fct{B}) yields a model with a single treatment factor, and this model reduction adheres again to the marginality principle (Fig.~\ref{fig:marginality}(d)). The resulting design resembles a one-way ANOVA and simplifies to a $t$-test situation with the familiar $2n-2$ error degrees of freedom if $a=2$.

\section{UNIT STRUCTURES: BLOCKING AND REPLICATION}
\label{sec:unitstructure}
While treatment structures are predominantly based on crossed factors, both crossing and nesting lead to commonly used unit structures. Multiple unit factors provide different possibilities for randomization, and identical unit- and treatment structures then lead to different experimental designs. In addition, unit-by-treatment interactions emerge and need to be addressed in the analysis or by treating them as negligible and removing them from the diagrams.  

\subsection{Nested Unit Factors}
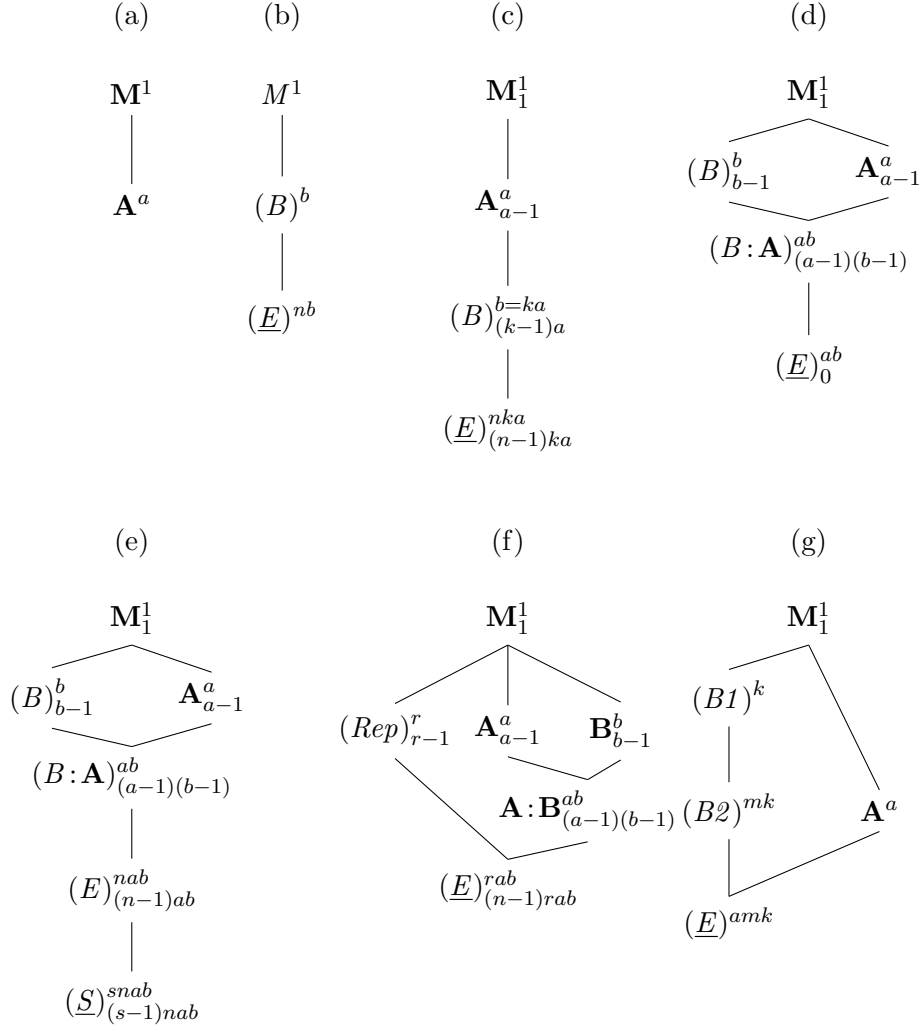
\begin{figure}
\centering
\usetikzlibrary{positioning}
\begin{tikzpicture}[
  var/.style={rectangle, draw=none, minimum size=5mm},
  node distance=1.5cm and 2cm,
  font=\small
  ]

    \node (ut)   {$\mathbf{M}^1$};
    \node (nitrot) [below of=ut] {$\mathbf{A}^a$};
    \draw (ut.south) -- (nitrot.north);
    \node (labelt) [above of=ut, node distance=1cm] {(a)};

    \node (uu) [right of= ut, node distance = 2cm] {$\mathit{M}^1$};
    \node (plotu) [below of=uu] {$\mathit{(B)}^b$};
    \node (subu) [below of=plotu] {$\mathit{(\underline{E})^{nb}}$};
    \draw (uu.south) -- (plotu.north);
    \draw (plotu.south) -- (subu.north);
    \node (labelu) [above of=uu, node distance=1cm] {(b)};
    
    \node (us) [right of=uu, node distance=3cm] {$\mathbf{M}^1_1$};
    \node (nitros) [below of=us] {$\mathbf{A}^a_{a-1}$};
    \node (plots) [below of=nitros] {$\mathit{(B)}^{b=ka}_{(k-1)a}$};
    \node (subs) [below of=plots] {$\mathit{(\underline{E})}^{nka}_{(n-1)ka}$};
    \draw (us.south) -- (nitros.north);
    \draw (nitros.south) -- (plots.north);
    \draw (plots.south) -- (subs.north);
    \node (labels) [above of=us, node distance=1cm] {(c)};
    
    \node (ur) [right of=us, node distance=4cm] {$\mathbf{M}^1_1$};
    \node (plotr) [below left of=ur] {$\mathit{(B)}^b_{b-1}$};
    \node (nitror) [below right of=ur] {$\mathbf{A}^a_{a-1}$};
    \node (plotnitror) [below left of=nitror] {$(\mathit{B}\!:\!\mathbf{A})^{ab}_{(a-1)(b-1)}$};
    \node (subr) [below of=plotnitror] {$\mathit{(\underline{E})}^{ab}_{0}$};
    \draw (ur.south) -- (plotr.north);
    \draw (ur.south) -- (nitror.north);
    \draw (plotr.south) -- (plotnitror.north);
    \draw (nitror.south) -- (plotnitror.north);
    \draw (plotnitror.south) -- (subr.north);
    \node (labelr) [above of=ur, node distance=1cm] {(d)};
 
    \node (ux) [below of=ut, node distance=7cm] {$\mathbf{M}^1_1$};
    \node (plotx) [below left of=ux] {$\mathit{(B)}^b_{b-1}$};
    \node (nitrox) [below right of=ux] {$\mathbf{A}^a_{a-1}$};
    \node (plotnitrox) [below left of=nitrox] {$(\mathit{B}\!:\!\mathbf{A})^{ab}_{(a-1)(b-1)}$};
    \node (subx) [below of=plotnitrox] {$\mathit{(E)}^{nab}_{(n-1)ab}$};
    \node (smpx) [below of=subx] {$\mathit{(\underline{S})}^{snab}_{(s-1)nab}$};
    \draw (ux.south) -- (plotx.north);
    \draw (ux.south) -- (nitrox.north);
    \draw (plotx.south) -- (plotnitrox.north);
    \draw (nitrox.south) -- (plotnitrox.north);
    \draw (plotnitrox.south) -- (subx.north);
    \draw (subx.south) -- (smpx.north);
    \node (labelx) [above of=ux, node distance=1cm] {(e)};
    
    \node (u) [below of=us, node distance=7cm] {$\mathbf{M}^1_1$};
    \node (a) [below of=u] {$\mathbf{A}^a_{a-1}$};
    \node (rep) [left of=a] {$\mathit{(Rep)}^r_{r-1}$};
    \node (b) [right of=a] {$\mathbf{B}^b_{b-1}$};
    \node (ab) [below right of=a] {$\mathbf{A\!:\!B}^{ab}_{(a-1)(b-1)}$};
    \node (e) [below left of=ab] {$(\underline{\mathit{E}})^{rab}_{(n-1)rab}$};
    \draw (u.south) -- (a.north);
    \draw (u.south) -- (b.north);
    \draw (u.south) -- (rep.north);
    \draw (a.south) -- (ab.north);
    \draw (b.south) -- (ab.north);
    \draw (ab.south) -- (e.north);        
    \draw (rep.south) -- (e.north);  
    \node (label) [above of=u, node distance=1cm] {(f)};
    
    \node (ub) [below of=ur, node distance=7cm] {$\mathbf{M}^1_1$};
    \node (b1b) [below left of=ub] {$\mathit{(B1)}^k$};
    \node (b2b) [below of=b1b] {$\mathit{(B2)}^{mk}$};
    \node (ab) [right of=b2b, node distance=2cm] {$\mathbf{A}^a$};
    \node (eb) [below of=b2b] {$(\underline{\mathit{E}})^{amk}$};
    \draw (ub.south) -- (b1b.north);
    \draw (ub.south) -- (ab.north);
    \draw (b1b.south) -- (b2b.north);
    \draw (ab.south) -- (eb.north);
    \draw (b2b.south) -- (eb.north);
    \node (label) [above of=ub, node distance=1cm] {(g)};
    
\end{tikzpicture}
\caption{(a): one-factor treatment structure. (b): unit structure with two nested factors. (c): \fct{A} randomized on \fct{B} with responses measured on \fct{E} gives a design with sub-sampling. (d): randomized complete block design (RCBD) with \fct{A} randomized on \fct{E} and crossed with \fct{B}. (e): generalized RCBD with treatment levels replicated per block and multiple samples per experimental unit. (f): replicating a treatment design by crossing with a unit factor. (g): design with two nested blocks resembles an RCBD with additional block structure.}
\label{fig:rcbd}
\end{figure}

We first consider a simple treatment structure with a single factor \fct{A} with $a$ levels and combine it with a unit structure with two nested factors \fct{B} ($b$ levels) and \fct{E} ($nb$ levels), with responses recorded for the lower factor \fct{E} (Fig.~\ref{fig:rcbd}(a,b)). This situation might correspond to an agricultural experiment with subplots nested in plots or to an animal experiment with mice nested in litters, for example. We now have two options for constructing the experiment design by randomizing the treatment factor on either the upper or lower unit factor. 

\subsubsection{Sub-sampling and Decomposing Variation}
\label{sub:subsampling}
If we randomize the treatment factor \fct{A} on the upper unit factor \fct{B} as in Figure~\ref{fig:rcbd}(c), then \fct{B} is the experimental unit for \fct{A} with $b=ka$ levels and provides the denominator for the $F$-test with $(a-1)$ and $(k-1)a$ degrees of freedom. Since \fct{E} is still the response unit, individual measurements are thus pseudo-replications~\citep{Hurlbert1984}.

The ANOVA table for this design has two error strata. If  $k>1$ and $n>1$, we have replication on both unit factors and can estimate their two variance components $\sigma^2_B$ and $\sigma^2_E$. The relevant variance for testing treatment effects is proportional to $\sigma_B^2+\sigma_E^2/n$ and the design is helpful only if the $\sigma_E^2$ is not small compared to $\sigma_B^2$, and thus precision can be gained by averaging several measurements for each experimental unit.

In \texttt{R} notation for ANOVA, the right-hand side of the model specification is \texttt{A+Error(B)} and we note that the term outside \texttt{Error()} corresponds to the treatment structure (with the general mean omitted as usual) while the term inside \texttt{Error()} corresponds to the unit structure (with the lowest factor omitted as usual). This observation simplifies the model specification in the frequent case that all fixed factors are in the treatment structure and all random factors are in the unit structure and there is no interaction between them. For a linear mixed model using the \texttt{lme4} package, the right-hand side is \fct{A+(1|B)} and using the Kenward-Roger approximation~\citep{Roger1997} ensures agreement between model and diagram degrees of freedom.  

The chain of unit factors can be elongated, for example, to separate variation of the experimental material, samples, individual sample preparations, and individual measurements. The contributions of each factor can only be estimated if the factor is replicated and are otherwise completely confounded with the next-higher factor in the chain. If no replication is present in the chain, then the overall variation is conceptually divided into individual contributions, but the analysis has to be based on a single estimate of the overall variation. Thus, it becomes clear that the lowest unit factor in any design contains all variation not explicitly captured by other factors.

\subsubsection{(Generalized) Randomized Complete Block Designs}
If we randomize the treatment factor on the lower unit factor \fct{E} as in Figure~\ref{fig:rcbd}(d), we automatically cross \fct{A} with \fct{B} and achieve a block design. Crossing a unit and a treatment factor imposes restrictions on the randomization of treatment factor levels on experimental units, which we need to account for in the analysis.

The two crossed factors give rise to a unit-by-treatment interaction factor \fct{A:B}. It only occurs in the experiment design diagram and is not present in either treatment or unit structure; it is a random factor because one of its constituent factors is random. The interaction is mathematically very similar to a treatment-by-treatment interaction, but its interpretation can be very different~\citep{cox1984,berringtondegonzalez2007}. 

Following the diagram, the treatment factor is tested against the unit-by-treatment interaction. If each treatment level occurs once per block, we derive the standard randomized complete block design (RCBD) shown in Figure~\ref{fig:rcbd}(d) with zero residual degrees of freedom. The treatment factor can only be tested if we assume that the unit-by-treatment interaction is negligible and we pool the interaction and the residual factor \fct{E} to provide the error mean squares. Because we often have substantial freedom in choosing a blocking factor, subject-matter knowledge should ensure the no-interaction assumption is reasonable. 

We can extend the design to a generalized randomized complete block design (GRCBD) by allowing multiple experimental units per block to receive the same treatment and/or by sampling each experimental unit multiple times~\citep{addelman1969}. The full diagram is shown in Figure~\ref{fig:rcbd}(e), where \fct{E} represents the experimental units and \fct{S} represents multiple samples per unit. We re-derive the RCBD with $n=s=1$. For $s=1, n>1$, we get a design where multiple units per block receive the same treatment, and we have one response per unit. For $s>1, n=1$, we get an RCBD, where each unit is measured multiple times. Again, a negligible unit-treatment interaction allows us to remove the factor \fct{A:B} from the diagram.

The (G)RCBD with non-negligible unit-by-treatment interaction cannot be analyzed using the standard \texttt{R} commands for ANOVA because it involves the interaction of a fixed and a random factor. An analysis of a GRCBD without sub-sampling, for example, then requires a linear mixed model with specification \fct{A+(1|B)+(1|A:B)}. Both RCBD and GRCBD also invite discussion of finding an $F$-test for the random block effect of \fct{B} and its possible interpretation~\citep{samuels1991}. Since the interaction factor involves \fct{A} as a new fixed factor, the correct error term for testing \fct{B} is provided by \fct{E}.

The idea of crossing a unit factor with a treatment factor can be extended to crossing it with the entire treatment structure. This yields a simple design pattern for replicating treatment designs. An example is shown in Figure~\ref{fig:rcbd}(f), where a two-factor factorial design is crossed with a unit factor \fct{Rep} to account for potential differences in replicates. For example, the logistics of the experiment might require performing replicates on multiple days, and calibration drift of measurement devices might lead to day-to-day differences. Assuming negligible interaction between replicate and treatments, we can remove the \fct{Rep:A:B} interaction from this design to yield an (G)RCBD. If \fct{Rep} additionally removes substantial variation from the experimental units \fct{E}, this strategy also yields efficient blocking.

\subsubsection{Blocks as Fixed Factors}
A different analysis emerges if we consider an (G)RCBD with a fixed rather than random block factor \fct{B}. Then, the analysis is similar to a two-factor completely randomized design with a single error stratum, with \fct{A},  \fct{B}, and \fct{A:B} now tested against \fct{E}, and the model is specified simply by \fct{A*B} including the interaction, and by \fct{A+B} for an additive model. The ANOVA table contains an $F$-test for the blocking factor \fct{B}, which invites discussion about its interpretation and comparison to a random blocking factor. An example would be testing different drug treatments on men and women, with sex considered as a fixed unit factor with two levels, and \fct{E} providing replication within each sex. In the notion of~\citet{cox1984}, \fct{B} is then an \emph{intrinsic} rather than a \emph{non-specific} unit factor, and the unit-by-treatment interaction might be of direct relevance to the interpretation of the experimental results and usually cannot be ignored.

\subsubsection{Nested Blocks}
The idea of the RCBD can be expanded in another direction, by considering several nested blocking factors, all crossed with the treatment factor. Such a design is shown in Figure~\ref{fig:rcbd}(g) for two nested blocking factors \fct{B1} and \fct{B2}, with $k$ and $mk$ levels, respectively. Again, we assume all unit-by-treatment interactions to be negligible. 

For treatment comparisons, this design effectively looks like a RCBD with a single blocking factor with $mk$ levels, but both factors contribute to the reduction in residual variance, and both enforce restrictions on the assignment of treatment factor levels to experimental units. The two blocking contributions can be separately estimated if $m>1$ and are confounded for $m=1$. For example, in an animal experiment, the levels of \fct{B1} might be cages, while levels of \fct{B2} are litters and the unit structure describes using $m$ litters per cage. The experiment might require keeping only one litter per cage, which confounds the litter-to-litter and cage-to-cage variations. The model is specified as \fct{A+Error(B1/B2)} for a classical ANOVA, and as \fct{A+(1|B1/B2)} for a mixed model analysis.

\subsection{Crossed Unit Factors}
Unit factors can also be crossed to allow simultaneous control for multiple sources of variation. We consider two crossed unit factors called \fct{R} (for row) and \fct{C} (for column) to emphasize the rectangular row-column form of the blocking. Each intersection of a row and column (a cell) corresponds to a level of the interaction \fct{R:C}, and several units might be nested in each cell. We consider a single treatment factor \fct{A} with $a$ levels, but more complex treatment structures are possible.

\subsubsection{All Treatments per Cell}
\begin{figure}[htbp]
\begin{center}
\includegraphics[width=0.5\textwidth]{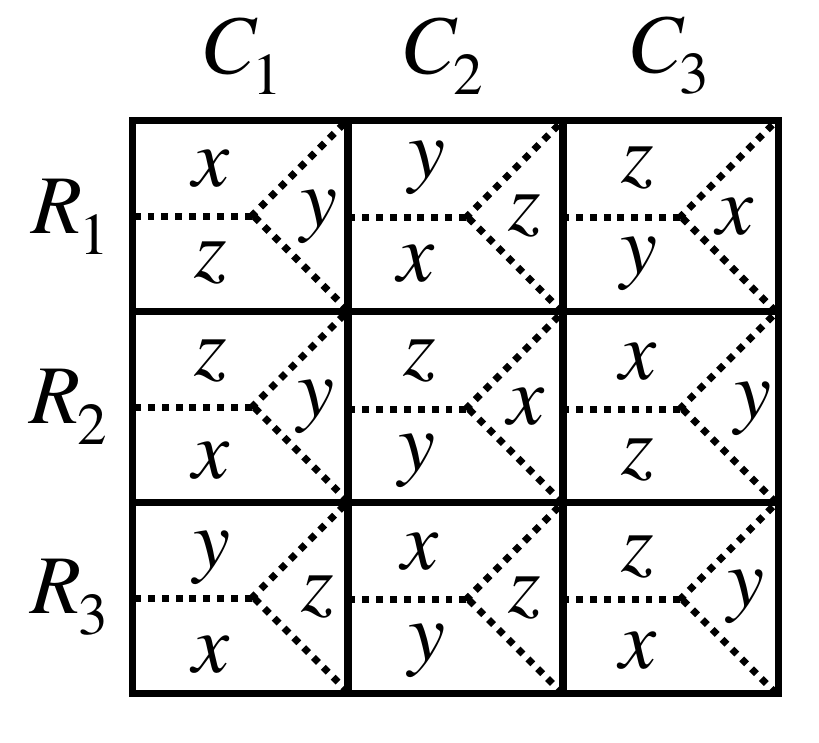}
\caption{Randomized data layout for design in Figure~\ref{fig:nestedblocks}(a) with $r=c=t=3$ and $n=1$. Treatments are crossed with interaction of row and column block factor, yielding independent randomization of treatment levels to units within row-column intersections.}
\label{fig:datalayout1}
\end{center}
\end{figure}

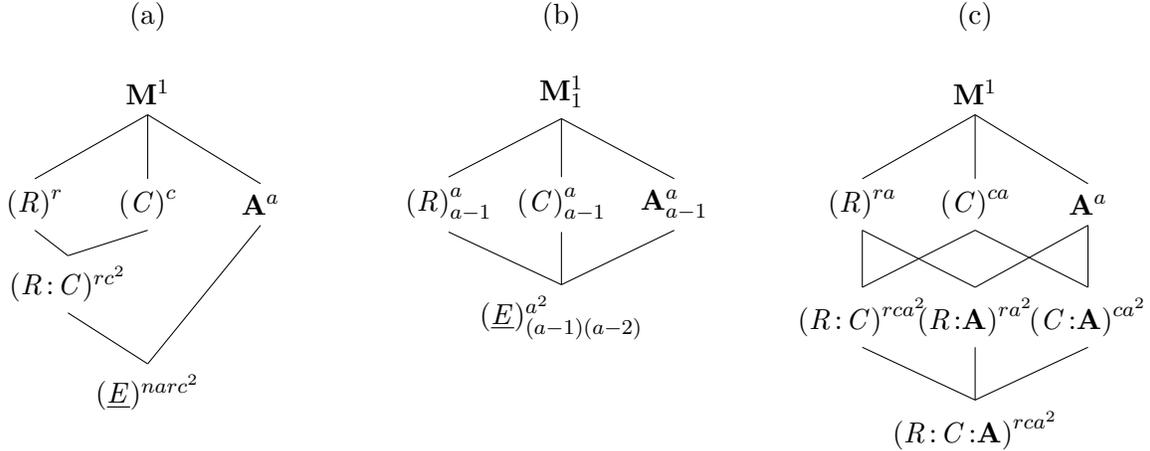
\begin{figure}
\centering
\usetikzlibrary{positioning}
\begin{tikzpicture}[
  var/.style={rectangle, draw=none, minimum size=5mm},
  node distance=1.5cm and 2cm,
  font=\small
  ]

    \node (ul) {$\mathbf{M}^1$};
    \node (cl) [below of=ul] {$\mathit{(C)}^{c}$};
    \node (rl) [left of=cl] {$\mathit{(R)}^{r}$};
    \node (al) [right of=cl] {$\mathbf{A}^a$};
    \node (rcl) [below left of=cl] {$(\mathit{R\!:\!C})^{rc^2}$};
    \node (el) [below of=cl, node distance=2.5cm] {$(\underline{\mathit{E}})^{narc^2}$};
    \draw (ul.south) -- (rl.north);
    \draw (ul.south) -- (cl.north);
    \draw (ul.south) -- (al.north);
    \draw (rl.south) -- (rcl.north);
    \draw (cl.south) -- (rcl.north);
    \draw (rcl.south) -- (el.north);  
    \draw (al.south) -- (el.north);  
    \node (labell) [above of=ul, node distance=1cm] {(a)};

    \node (us) [right of=ul, node distance=5.5cm] {$\mathbf{M}^1_1$};
    \node (cs) [below of=us] {$\mathit{(C)}^{a}_{a-1}$};
    \node (rs) [left of=cs] {$\mathit{(R)}^{a}_{a-1}$};
    \node (as) [right of=cs] {$\mathbf{A}^a_{a-1}$};
    \node (es) [below of=cs] {$(\underline{\mathit{E}})^{a^2}_{(a-1)(a-2)}$};
    \draw (us.south) -- (rs.north);
    \draw (us.south) -- (cs.north);
    \draw (us.south) -- (as.north);
    \draw (rs.south) -- (es.north);
    \draw (cs.south) -- (es.north);
    \draw (as.south) -- (es.north);  
    \node (labels) [above of=us, node distance=1cm] {(b)};

    \node (uf) [right of=us, node distance=5.5cm] {$\mathbf{M}^1$};
    \node (cf) [below of=uf] {$\mathit{(C)}^{ca}$};
    \node (rf) [left of=cf] {$\mathit{(R)}^{ra}$};
    \node (af) [right of=cf] {$\mathbf{A}^a$};
    \node (rcf) [below of=rf] {$\mathit{(R\!:\!C)}^{rca^2}$};
    \node (raf) [below of=cf] {$\mathit{(R\!:}\mathbf{A}\mathit{)}^{ra^2}$};
    \node (caf) [below of=af] {$\mathit{(C\!:}\mathbf{A}\mathit{)}^{ca^2}$};
    \node (rcaf) [below of=raf] {$\mathit{(R\!:\!C\!:}\mathbf{A}\mathit{)}^{rca^2}$}; 
    \draw (uf.south) -- (rf.north);
    \draw (uf.south) -- (cf.north);
    \draw (uf.south) -- (af.north);
    \draw (rf.south) -- (rcf.north);
    \draw (rf.south) -- (raf.north);
    \draw (cf.south) -- (rcf.north);
    \draw (cf.south) -- (caf.north);
    \draw (af.south) -- (caf.north);
    \draw (af.south) -- (raf.north);
    \draw (raf.south) -- (rcaf.north);
    \draw (caf.south) -- (rcaf.north);
    \draw (rcf.south) -- (rcaf.north);
    \node (labelf) [above of=uf, node distance=1cm] {(c)};    
    
\end{tikzpicture}
\caption{(a): two crossed unit factors provide a block with internal structure for the treatment factor. (b): the classical Latin square design. (c): with non-negligible block-by-treatment interactions, no exact $F$-test exists for the treatment factor.}
\label{fig:nestedblocks}
\end{figure}

A first design uses $r$ rows and $c$ columns and crosses their interaction with a simple treatment structure as shown in Figure~\ref{fig:nestedblocks}(a), with $n$ replicates of each treatment level per cell. For simplicity, we again assume that unit-by-treatment interactions are negligible; the model is then given by \fct{A+Error(R*C)} and \fct{A+(1|R)+(1|C)+(1|R:C)}, respectively. The corresponding data layout is shown in Figure~\ref{fig:datalayout1}.

The rectangular blocking appears as a simple blocking with $narc$ levels, and the design is thus similar to a (G)RCBD, but imposes additional structure in the blocking and restrictions on the treatment randomization.

A different design emerges if we randomize the treatments directly on the interaction factor, such that each cell gets assigned a single treatment level. A balanced assignment requires that $r$ and $c$ are multiples of $a$, and pseudo-replication occurs for $n>1$. An important example is the Latin square shown in Figure~\ref{fig:nestedblocks}(b) which uses $r=c=a$ and $n=1$, and thus has $a^2$ experimental units, such that each treatment level occurs exactly once per row and column. The factor \fct{E} in Figure~\ref{fig:nestedblocks}(b) then provides the error term for testing the treatment factor and all two-way interactions have to be assumed negligible in this design.

An interesting situation occurs if we cannot assume that two-way interactions are negligible, as shown in Figure~\ref{fig:nestedblocks}(c). Following edges from \fct{A} downwards, we find that there is no unique factor for providing the denominator mean squares for an $F$-test of the treatment. Indeed, as is well known, no exact $F$-test exists for this situation, and we need to resort to finding an approximation of the denominator mean squares and degrees of freedom, such as the weighted averages given by \citet{satterthwaite1946}. Note again that the two-way interactions are qualitatively different: while \fct{R:C} is a combination of blocking levels (corresponding to a cell), both \fct{R:A} and \fct{C:A} are unit-treatment interactions.

\subsubsection{Replicating Latin Square Designs}
While the Latin square is the most common design based on two crossed unit factors, it might suffer from insufficient error degrees of freedom: for $a=2$, $a=3$, and $a=4$ treatment levels, these are $0$, $2$, and $6$, respectively. It is therefore often necessary to replicate a Latin square design and we illustrate the use of Hasse diagrams for discussing different strategies for this replication.  In each case, the Hasse diagram allows us to quickly calculate the resulting error degrees of freedom and to compare them between the designs. Moreover, the specification of an appropriate linear model is again quickly derived from the diagrams.

First, we might consider replicating rows while keeping the columns, using $ra$ row factor levels instead of $a$, for example. If we do not impose any restrictions on the assignment of treatments to rows other than overall balancedness, consecutive sets of $a$ rows do not need to form Latin squares, and we arrive at a design known as \emph{Latin rectangle}. It corresponds to Figure~\ref{fig:nestedblocks}(a) with $c=1$ and negligible interaction \fct{R:C}; the  data layout for this design is shown in Figure~\ref{fig:datalayout2}(a). The linear model is thus given in \texttt{R} notation by \fct{A+Error(R+C)} for an ANOVA, and \fct{A+(1|R)+(1|C)} for a mixed model.

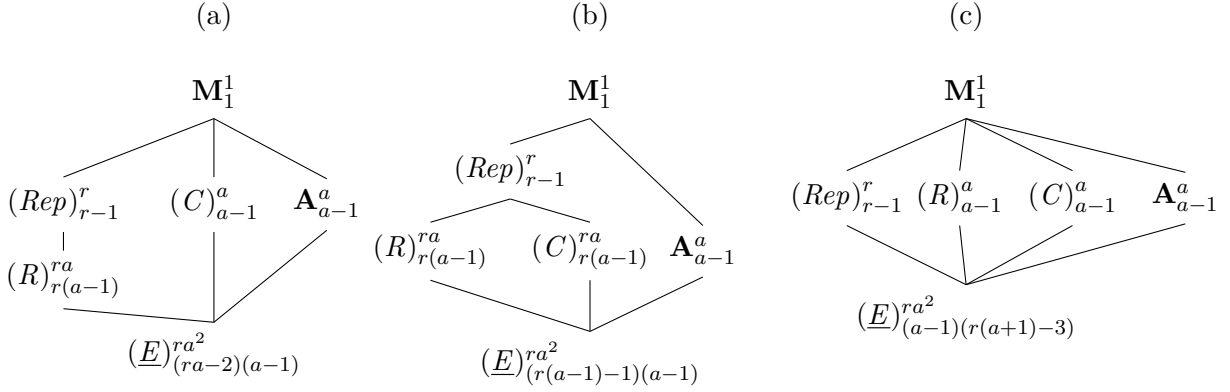
\begin{figure}
\centering
\usetikzlibrary{positioning}
\begin{tikzpicture}[
  var/.style={rectangle, draw=none, minimum size=5mm},
  node distance=1.5cm and 2cm,
  font=\small
  ]

    \node (ul) {$\mathbf{M}^1_1$};
    \node (cl) [below of=ul] {$\mathit{(C)}^{a}_{a-1}$};
    \node (repl) [left of=cl, node distance=2cm] {$\mathit{(Rep)}^{r}_{r-1}$};
    \node (rl) [below of=repl, node distance=1cm] {$\mathit{(R)}^{ra}_{r(a-1)}$};
    \node (al) [right of=cl] {$\mathbf{A}^a_{a-1}$};
    \node (el) [below of=cl, node distance=2cm] {$(\underline{\mathit{E}})^{ra^2}_{(ra-2)(a-1)}$};
    \draw (ul.south) -- (repl.north);
    \draw (repl.south) -- (rl.north);
    \draw (ul.south) -- (cl.north);
    \draw (ul.south) -- (al.north);
    \draw (rl.south) -- (el.north);
    \draw (cl.south) -- (el.north);
    \draw (al.south) -- (el.north);  
    \node (labell) [above of=ul, node distance=1cm] {(a)};
        
    \node (ur) [right of=ul, node distance=5cm] {$\mathbf{M}^1_1$};
    \node (repr) [below left of=ur] {$\mathit{(Rep)}^{r}_{r-1}$};
    \node (rr) [below left of=repr] {$\mathit{(R)}^{ra}_{r(a-1)}$};
    \node (cr) [below right of=repr] {$\mathit{(C)}^{ra}_{r(a-1)}$};
    \node (ar) [right of=cr] {$\mathbf{A}^a_{a-1}$};
    \node (er) [below of=cr] {$(\underline{\mathit{E}})^{ra^2}_{(r(a-1)-1)(a-1)}$};
    \draw (ur.south) -- (repr.north);
    \draw (repr.south) -- (cr.north);
    \draw (repr.south) -- (rr.north);
    \draw (ur.south) -- (ar.north);
    \draw (rr.south) -- (er.north);
    \draw (cr.south) -- (er.north);
    \draw (ar.south) -- (er.north); 
    \node (labelr) [above of=ur, node distance=1cm] {(b)};    
    
    \node (uf) [right of=ur, node distance=5cm] {$\mathbf{M}^1_1$};
    \node (cf) [below right of=uf, node distance=2cm] {$\mathit{(C)}^{a}_{a-1}$};
    \node (rf) [left of=cf] {$\mathit{(R)}^{a}_{a-1}$};
    \node (repf) [left of=rf] {$\mathit{(Rep)}^{r}_{r-1}$};
    \node (af) [right of=cf] {$\mathbf{A}^a_{a-1}$};
    \node (ef) [below of=uf] [node distance=3cm] {$(\underline{\mathit{E}})^{ra^2}_{(a-1)(r(a+1)-3)}$};
    \draw (uf.south) -- (repf.north);
    \draw (uf.south) -- (rf.north);
    \draw (uf.south) -- (cf.north);
    \draw (uf.south) -- (af.north);
    \draw (repf.south) -- (ef.north);
    \draw (rf.south) -- (ef.north);
    \draw (cf.south) -- (ef.north);
    \draw (af.south) -- (ef.north); 
    \node (labelf) [above of=uf, node distance=1cm] {(c)};
        
\end{tikzpicture}
\caption{Replication of Latin square design. (a): $r$ sets of $a$ rows, and identical columns per replicate. (b): new rows and columns in each replicate. (c): identical rows and columns in each replicate.}
\label{fig:lsrep}
\end{figure}

\begin{figure}[htbp]
\begin{center}
\includegraphics[width=\textwidth]{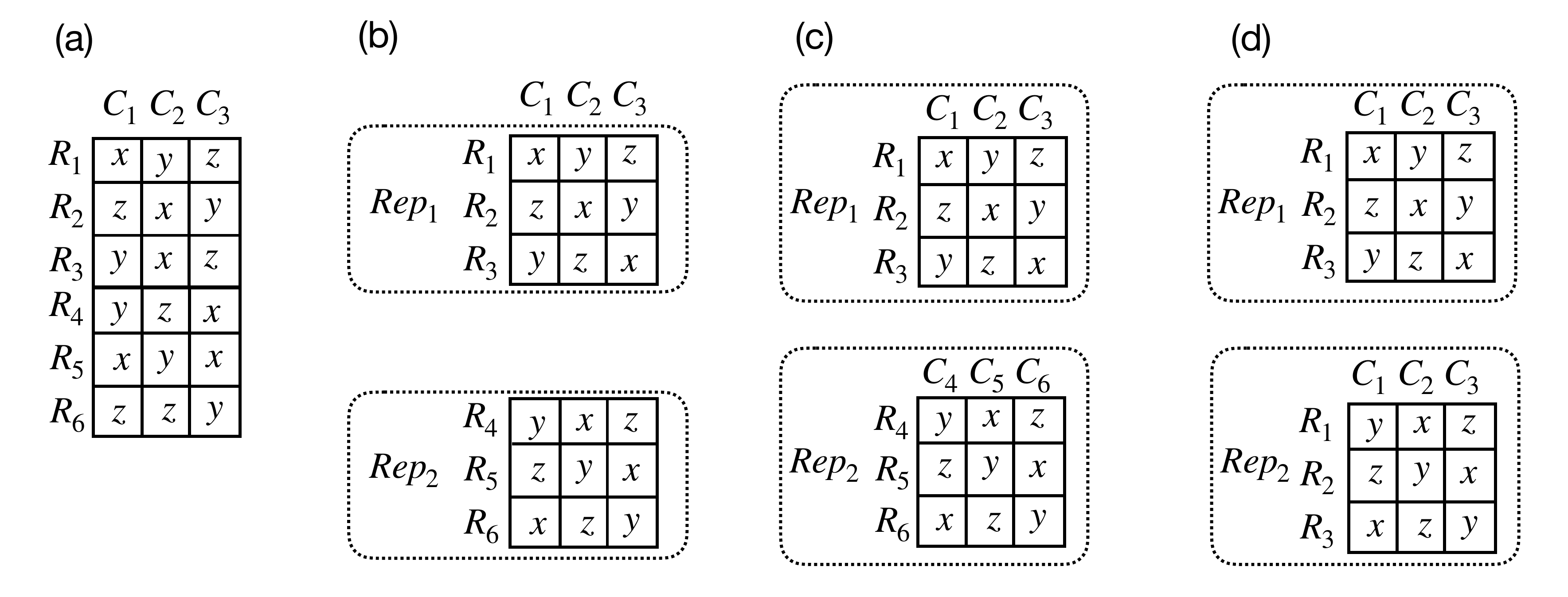}
\caption{Data layout for replicating Latin square designs. (a): Latin rectangle with two replicates of three rows; replicates do not form Latin squares. (b): two independent replicates of rows and identical columns, each replicate is a Latin square (cf. diagram Fig.~\ref{fig:lsrep}(a)). (c): two independent replicates of rows and columns, each replicate is a Latin square (cf. diagram Fig.~\ref{fig:lsrep}(b)). (d): two replicates of Latin square with identical rows and columns, but independent randomization of treatments (cf. diagram Fig.~\ref{fig:lsrep}(c)).}
\label{fig:datalayout2}
\end{center}
\end{figure}

Another option is using $r$ replicates of $a$ rows again, but to additionally require that each replicate forms an independent Latin square. We achieve this by introducing a new unit factor \fct{Rep} to account for the replication and arrive at the design shown in Figure~\ref{fig:lsrep}(a) with \fct{R} nested in \fct{Rep}; the data layout is shown in Figure~\ref{fig:datalayout2}(b). Since the treatment factor \fct{A} is crossed with the new unit factor \fct{Rep}, each treatment level must occur once in each row per replicate, and this design corresponds to $r$ independently randomized Latin squares, where the column factor levels are identical in each replicate, but the row factor levels are different. The model specifications are \fct{A+Error(Rep/R+C)} (ANOVA) and \fct{A+(1|Rep/R)+(1|C)} (LMM), respectively.

By nesting both \fct{R} and \fct{C} in \fct{Rep}, we find yet another useful design that uses different row and column factor levels in each replicate of the Latin square (cf.\ diagram in Fig.~\ref{fig:lsrep}(b) and data layout in Fig.~\ref{fig:datalayout2}(c)) with models \fct{A+Error(Rep/(R+C))} (ANOVA) and \fct{A+(1|Rep)+(1|Rep:R)+(1|Rep:C)} (LMM). Finally, by crossing \fct{Rep} with the row and column factor we keep the same rows and columns in each replicate and re-randomize the treatment levels independently on each square. The diagram is shown in Figure~\ref{fig:lsrep}(c) and the data layout in Figure~\ref{fig:datalayout2}(d) with models \fct{A+Error(Rep+R+C)} (ANOVA) and \fct{A+(1|Rep)+(1|R)+(1|C)} (LMM).
  
\section{SPLIT-UNIT DESIGNS}
\label{sec:splitunit}
\begin{figure}
\centering
\begin{tikzpicture}[
  var/.style={rectangle, draw=none, minimum size=5mm},
  node distance=1.5cm and 2cm,
  font=\small
  ]

    \node (ut)   {$\mathbf{U}^1$};
    \node (vart) [below left of=ut] {$\mathbf{Variety}^3$};
    \node (nitrot) [below right of=ut] {$\mathbf{Nitrogen}^4$};
    \node (varnitrot) [below right of=vart] {$\mathbf{Variety\!:\!Nitrogen}^{12}$};
    \draw (ut.south east) -- (nitrot.north);
    \draw (ut.south west) -- (vart.north);
    \draw (nitrot.south) -- (varnitrot.north);
    \draw (vart.south) -- (varnitrot.north);
    \node (labelt) [above of=ut, node distance=1cm] {(a)};

    \node (uu) [right of= ut, node distance = 4cm] {$\mathit{U}^1$};
    \node (blocku) [below of=uu] {$\mathit{(Block)^6}$}; 
    \node (plotu) [below of=blocku] {$\mathit{(Plot)^{18}}$};
    \node (subu) [below of=plotu] {$\mathit{(\underline{Subplot})^{72}}$};
    \draw (uu.south) -- (blocku.north);
    \draw (blocku.south) -- (plotu.north);    
    \draw (plotu.south) -- (subu.north);
    \node (labelu) [above of=uu, node distance=1cm] {(b)};
    
    \node (u) [right of=uu, node distance=5cm] {$\mathbf{U}^1_1$};
    \node (var) [below of=u] {$\mathbf{Variety}^3_2$};
    \node (nitro) [right of=var, node distance=2cm] {$\mathbf{Nitrogen}^4_3$};
    \node (varnitro) [below of=nitro] {$\mathbf{Variety\!:\!Nitrogen}^{12}_6$};
    \node (block) [left of=var, node distance=2cm] {$\mathit{(Block)^{6}_5}$};
    \node (plot) [below of=block, node distance=3cm] {$\mathit{(Plot)}^{18}_{10}$};
    \node (subplot) [below of=plot] {$\mathit{(\underline{Subplot})}^{72}_{45}$};
    \draw (block.south) -- (plot.north);
    \draw (plot.south) -- (subplot.north);
    \draw (u.south west) -- (block.north);
    \draw (u.south) -- (var.north);
    \draw (u.south east) -- (nitro.north);
    \draw (nitro.south) -- (varnitro.north);
    \draw (var.south) -- (varnitro.north);
    \draw (var.south) -- (plot.north);
    \draw (varnitro.south) -- (subplot.north east);
    \node (label) [above of=u, node distance=1cm] {(c)};

\end{tikzpicture}
\caption{Hasse diagrams of a split-unit design with oat varieties randomized on plots, and nitrogen amounts to subplots within plots, the whole design replicated in blocks. (a): treatment structure. (b): unit structure. (c): experiment structure.}
\label{fig:splitunit}
\end{figure}
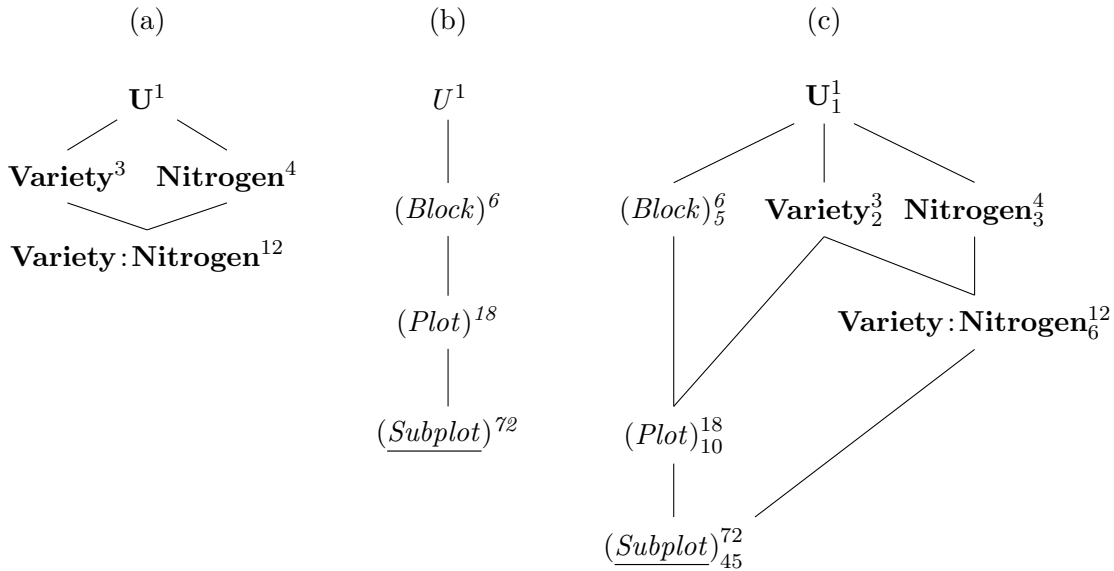

As a last illustration, we consider a classical agricultural example provided in~\citet{yates1935}, with three oat varieties tested at four nitrogen levels each. Oat varieties are randomized on larger plots of land, while the four nitrogen levels are randomized on subplots within each plot, and the experiment is replicated in six blocks. The diagrams for this example are given in Figure~\ref{fig:splitunit}. The treatment structure is a simple factorial design with two factors, and the unit structure consists of a simple chain of three nested factors. However, the treatment factors are now randomized on different unit factors, resulting in a split-unit design for the experiment structure. Note how some of the previous designs occur as part of the experiment diagram, including blocking a full treatment structure, sub-sampling, and crossing a treatment with nested blocks.

The diagram makes explicit that this design has two experimental unit factors with \fct{Plot} providing the experimental unit for \fct{Variety}, while \fct{Subplot} provides the experimental unit for both \fct{Nitrogen} and the interaction \fct{Variety:Nitrogen}. As a consequence, contrasts on \fct{Variety} main effects profit from variance reduction via \fct{Block}, but still encounter the combined variance of \fct{Plot} and \fct{Subplot} and the low replication on \fct{Plot}; the nested \fct{Subplot} resembles pseudo-replication. On the other hand, contrasts on \fct{Nitrogen} main effects and the interaction profit from reduced variance due to the blocking \fct{Plot} and \fct{Block} (resembling the design of Fig.~\ref{fig:nestedblocks}(a)) and from higher replication on \fct{Subplot}. 

In order to improve estimation and testing on the whole-unit factor \fct{Variety}, we need to improve on the blocking by \fct{Block} or increase the replication on \fct{Plot}. Both measures also positively impact estimation and testing involving the two other treatment factors. The linear model is specified by \fct{Variety*Nitrogen+Error(Block/Plot)} respectively by \fct{Variety*Nitrogen+(1|Block/Plot)}. The model specification again decomposes nicely into two very simple parts describing the treatment and the unit structures, respectively. A very common alternative specification is \fct{Variety*Nitrogen+Error(Variety/Nitrogen)} and exploits that the unit structure is implicit in combination of levels of the treatment factors. It is however an illogical model formula and a source of much confusion for students.

Using Hasse diagrams, it is straightforward to expand the idea of a split-unit design to more ``splits'', to designs such as split-blocks (or criss-cross), and to split-unit designs with more complex whole- or sub-unit designs, such as using a Latin square for the whole-unit factor.

\section{CONCLUSIONS}
\label{sec:conclusions}
The teaching of introductory experimental design to undergraduates and non-statisticians remains an important task for the statistical community, and it has been argued that such course provides additional benefits for students with comparatively little mathematical background~\citep{Blades2015}. In our own teaching of experimental design as second statistics course for biotechnology master students, we found that Hasse diagrams enable a more direct approach to establishing the classical ``good'' designs and allow recognition of design patterns shared between designs. They encourage free composition of designs from unit and treatment structures; the model specification for a resulting design can be derived directly from the experiment diagram and used with statistical software. The skeleton ANOVA table with factors, error strata, degrees of freedom, sum of squares, and expected mean squares can be derived directly from the experiment diagram~\citep{lohr2006}. This assists in teaching properties of a design and additionally allows students to check software output against a diagram to ensure compatibility of design and analysis.

While Hasse diagrams are suitable for many designs created by nesting and crossing factors, their main power lies with balanced and orthogonal designs. They are, however, not applicable for designs with more complex confounding, such as balanced incomplete block designs or Youden squares, but these designs can be developed once the RCBD and Latin squares are mastered. Hasse diagrams are also not well suited for explaining fractional factorial designs, although this could be done by introducing pseudo-factors as in~\citet{Bailey2008}. Again, these can be developed once full factorial designs are understood. 

We did not discuss more complex treatment structures such as higher-order factorial treatment designs, but their Hasse diagrams are easily derived. With more than two treatment factors, interactions of the same order are now crossed and give rise to new factors (namely the next-higher order interactions) in the diagram. This results in exponentially many factors in the treatment structure and methods for reducing this number include removal of higher-order interactions and fractions of the design. For the important special case of a $2^k$ factorial, the Hasse diagram immediately shows that each treatment factor has precisely one degree of freedom, and thus fractional factorials can be constructed by exploiting either the well-known additive $0/1$ or the multiplicative $-1/+1$ algebra to encode factor levels.

We also did not discuss random treatment factors. These are straightforward to capture in the diagrams, but necessarily require analysis by linear mixed models instead of a classical ANOVA. 

Some more recent developments extend the ideas of Hasse diagrams to larger classes of design problems. Although typically not relevant for introductory courses, they might stimulate similar uses of diagrams for more advanced courses and are certainly interesting for many consultation projects. Bate and co-workers~\citep{Bate2016, Bate2016a} propose diagrams that allow more complex response structures and are tailored towards linear mixed models with covariance structures beyond simple variance components. Independently, the work by Brien and co-workers extends the description from a single experiment to a chain of ``tiered'' experiments~\citep{Brien1983,Brien1999,Brien2009} which use multiple randomization. In reality, many experiments are of course multi-tiered, such as treatments randomly allocated to experimental units, and multiple measurement devices are then randomly assigned to record the response values. It is then easy to ``break'' the overall randomization when assigning all units with the same treatment level to the same measurement device, for example.

\bibliographystyle{agsm}
\bibliography{doe-hasse-references}
\end{document}